# THE INNER RINGS OF β PICTORIS

Z. Wahhaj,[1,2] D. W. Koerner,[2,3] M. E. Ressler,[3,4] M. W. Werner,[3,4] D. E. Backman,[3,5] and A. I. Sargent[3,6]

## ABSTRACT

We present Keck images of the dust disk around β Pictoris at $\lambda = 17.9$ μm that reveal new structure in its morphology. Within 1″ (19 AU) of the star, the long axis of the dust emission is rotated by more than 10° with respect to that of the overall disk. This angular offset is more pronounced than the warp detected at 3″.5 by the *Hubble Space Telescope* (*HST*) and is in the opposite direction. By contrast, the long axis of the emission contours ∼1″.5 from the star is aligned with the *HST* warp. Emission peaks between 1″.5 and 4″ from the star hint at the presence of rings similar to those observed in the outer disk at ∼25″ with the *HST* Space Telescope Imaging Spectrograph. A deconvolved image strongly suggests that the newly detected features arise from a system of four noncoplanar rings. Bayesian estimates based on the primary image lead to ring radii of $14 \pm 1$, $28 \pm 3$, $52 \pm 2$, and $82 \pm 2$ AU, with orbital inclinations that *alternate* in orientation relative to the overall disk and decrease in magnitude with increasing radius. We believe these new results make a strong case for the existence of a nascent planetary system around β Pic.

*Subject headings:* circumstellar matter — infrared: stars — planetary systems: formation —
planetary systems: protoplanetary disks — solar system: formation

## 1. INTRODUCTION

It has been nearly two decades since *IRAS* detections of circumstellar dust profoundly influenced our thinking about the prevalence and formation of extrasolar planetary systems (Aumann et al. 1984; Gillett 1986). Follow-up coronagraphic imaging of a disk around one *IRAS* source, β Pictoris, was particularly compelling (Smith & Terrile 1984) and contributed to its status as a prototype for optically thin "debris disks" around nearby main-sequence stars (Backman & Paresce 1993). Although these disks surrounded stars with ages of $10^7$–$10^8$ yr, the survival timescale of their constituent dust grains was estimated to be only $10^3$–$10^6$ yr and suggested a population of colliding planetesimals that replenished the population of dust grains. Debris disks apparently represent a derivative population of grains arising from planetesimal collisions rather than a remnant population from the original protostellar core. Unfolding properties of the disk around β Pic strongly supported this interpretation.

The star β Pic is a 20 Myr old (Barrado y Navascués et al. 1999) A5 V star at a distance of 19.28 pc from Earth (Crifo et al. 1997). Its disk has been imaged out to a radius of ∼1800 AU using optical coronagraphs (Smith & Terrile 1984; Golimowski, Durrance, & Clampin 1993; Kalas & Jewitt 1995; Larwood & Kalas 2001). Models of associated *IRAS* emission indicated that material was substantially depleted within 80 AU of the star (Backman, Gillett, & Witteborn 1992). Modeling of images at $\lambda = 10$ and 12 μm provided confirmation of this view (Lagage & Pantin 1994; Pantin, Lagage, & Artymowicz 1997). Asymmetries in brightness and morphology were discovered in images at optical, infrared, and submillimeter wavelengths (Kalas & Jewitt 1995; Pantin et al. 1997; Holland et al. 1998), and a 4° warp was detected by the *Hubble Space Telescope* (*HST*) within 4″of the star (Burrows et al. 1995; Heap et al. 2000). The gravitational influence of an orbiting substellar or planetary body was invoked to account for many of these features (e.g., Gillet 1986; Diner & Appleby 1986; Mouillet et al. 1997; Augereau et al. 2001).

Direct imaging of other debris disks was made possible by advances in high-resolution imaging at long wavelengths where the star-disk brightness contrast is greatly reduced. Submillimeter imaging with the James Clark Maxwell Telescope revealed asymmetric rings around Vega, α PsA, and ε Eri (Holland et al. 1998; Greaves et al. 1998). Mid-infrared array imaging at the Keck telescope provided views of 20 μm emission ostensibly from a narrow ring around HR 4796A (Koerner et al. 1998; Jayawardhana et al. 1998; Telesco et al. 2000). This discovery was dramatically confirmed with *HST* coronagraphic imaging at near-infrared wavelengths (Schneider et al. 1999) as was the detection of a disk around HD 141569A in which a prominent gap was also observed (Weinberger et al. 1999). Individual features have been explained by a variety of theoretical explanations (Kenyon et al. 1999; Takeuchi & Artymowicz 2001; Klahr & Lin 2001), but only the dynamical influence of substellar or planetary bodies has successfully explained them all (Takeuchi, Miyama, & Lin 1996; Bryden et al. 1999; Ozernoy et al. 2000; Lecavelier Des Etangs et al. 1996). Here we test the explanation more fully with a high-resolution imaging study at 17.9 μm of the inner region of the disk around β Pictoris.

## 2. OBSERVATIONS AND RESULTS

The star β Pictoris was observed at the f/40 bent Cassegrain focus of the Keck II telescope on UT dates 1998 March 14 and 15 and 1999 January 30 using MIRLIN (Mid-InfraRed Large-well Imager), a mid-infrared camera with a Boeing $128 \times 128$ pixel, high-flux Si:As blocked impurity band detector (Ressler et al. 1994). At Keck, MIRLIN has a plate scale of 0″.138 pixel$^{-1}$ and a 17″.5 field of view. A filter centered at $\lambda = 17.9$ μm with a width of 2 μm was used for the observations presented here. Sky subtraction was carried out by dif-ferencing images chopped 8″ in an east-west direction at a frequency of 4 Hz. The telescope was nodded 20″ in a north-south direction after co-adding several hundred frames at each of the chop positions. The total on-source integration time was 30 minutes. Disk emission in each of the double-differenced chop and nod frames was shifted and added to produce a final combined image. Observation and data reduction of standards α Car, β Gem, and α Tau proceeded in the same way for calibration purposes.

[1] University of Pennsylvania, David Rittenhouse Laboratory, 209 South 33d Street, Philadelphia, PA 19104-6396.
[2] Northern Arizona University, Department of Physics and Astronomy, Building 19, Room 209, Flagstaff, AZ 86001-6010.
[3] Visiting Astronomer, W. M. Keck Observatory, 65-1120 Mamalahoa Highway, Kamuela, HI 96743.
[4] Jet Propulsion Laboratory, California Institute of Technology, 4800 Oak Grove Drive, Pasadena, CA 91109.
[5] Physics and Astronomy Department, Franklin and Marshall College, P.O. Box 3003, Lancaster, PA 17604.
[6] Division of Physics, Mathematics, and Astronomy, MS 103-33, California Institute of Technology, Pasadena, CA 91125.

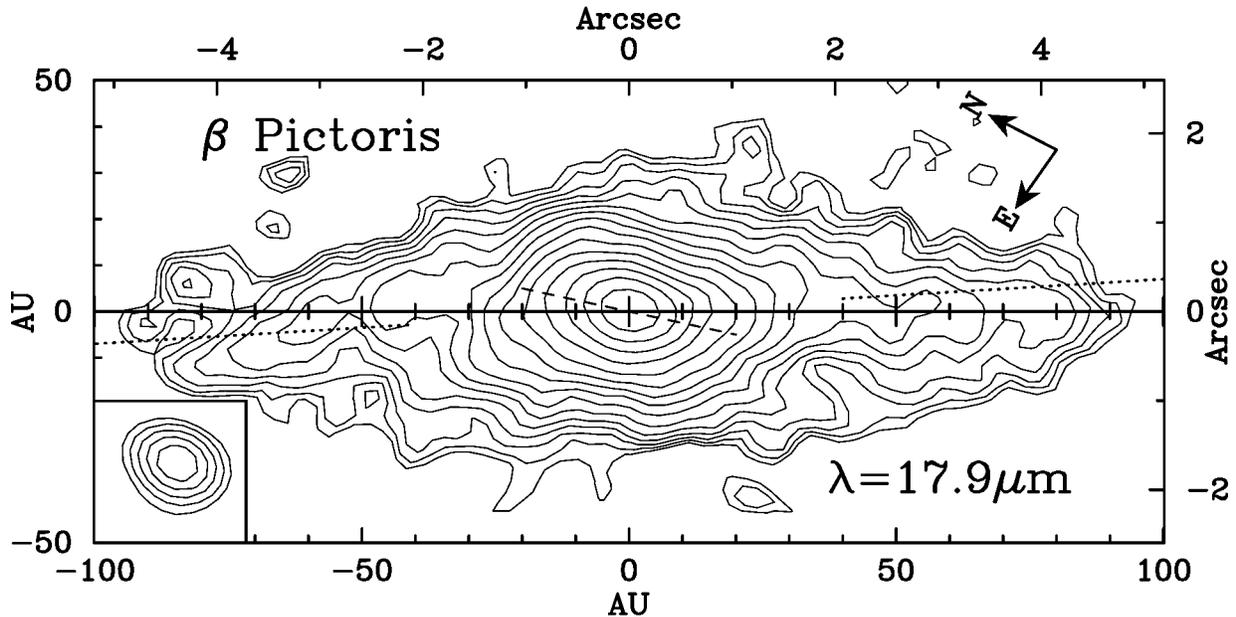

Fig. 1.—The β Pic Keck/MIRLIN image at 17.9 μm obtained by combining observations from three separate nights. The image has been rotated 59°.2 counterclockwise so that the outermost disk midplane at P.A. = 30°.8 is aligned with the horizontal axis. The background flux has an rms value of 19.2 mJy arcsec$^{-2}$, and the peak flux is 743 mJy arcsec$^{-2}$. The lowest contour is at the 2 σ level, and the $n$th contour is at 2 σ ($10^{0.087n}$). The dashed line shows the orientation of core emission (14° offset from the disk midplane). The dotted line shows the orientation of the STIS warp (4° offset).

Images from each of the three observing nights were smoothed with a circular hat function with radius equal to the half-maximum radius of the point-spread function (PSF) as measured in imaging of α Car taken immediately afterward. Smoothed images of β Pic were added after weighting by the signal-to-noise ratio of their peak flux. The result is displayed in the contour image in Figure 1, where the $x$-axis is aligned with the long axis of the outer disk at a position angle (P.A.) of 30°.8 (Kalas et al. 1995). PSFs from each night were combined in a similar fashion, yielding a PSF FWHM of 0″.7. The result is displayed in the inset to Figure 1. For β Pic, the total flux density in a 10″ diameter circular area centered on the star is 3.9 ± 0.8 Jy, nearly 8 times the value of 490 mJy expected based on extrapolation from optical and near-infrared photometry (Koornneef 1983) with the aid of a model of a photosphere with $T_{eff}$ = 8250 K (Kurucz 1993). IRAS fluxes at 12 and 25 μm are 3.5 and 9 Jy, respectively, consistent with the present measurement. The peak surface brightness is 743 mJy arcsec$^{-2}$ for a region bounded by the half-maximum radius of the PSF (0″.35). This is nearly 3 times the value of 280.5 mJy arcsec$^{-2}$ expected from the stellar photosphere in the same area. We conclude that warm dust grains are the dominant source of the emission seen in Figure 1.

The axis of core emission in Figure 1 deviates from the outer disk plane in the *opposite sense* to the warp observed with *HST* at optical wavelengths (Burrows et al. 1995). The latter is oriented along P.A. = 35°, an offset of −4°, and extends from 1″.5 (30 AU) to 4″.7 (90 AU) on both sides of the star (Heap et al. 2000). For comparison, the trend of the *HST* warp is marked with a dotted line in Figure 1. A first estimate of the degree of misalignment of the inner contours in Figure 1 was obtained by fitting elliptical Gaussians to the core of emission using NRAO's Astronomical Image Processing System task JMFIT. Fits were obtained inside boxed regions centered on the star and ranging in size from 1″.1 to 2″.8 on a side. The P.A. of the long axis shifted continuously from 17° to 29°, as increasingly large boxes were used, corresponding to a gradual shift from 14° to 2° with respect to the outer disk orientation, in excellent agreement with the 13° ± 2° measured by Weinberger et al. (2002) using the Long Wavelength Spectrometer on Keck I. Gaussian fits to the core of emission in images of α Car had axis ratios closer to unity and disparate orientations of the long-axis P.A., arguing strongly against an artifactual origin for the offset of emission. Indeed, we were unable to simulate the offset by convolution of an unwarped model disk with images of α Car.

A two-component Gaussian fit to the inner emission gives a more inclined angle for the inner component, ∼18°, and a second component with an orientation offset by only a few degrees from the disk plane in the direction of the warp seen with *HST*. Farther from the star, disk emission is clumped with peaks centered at stellocentric distances of 2″.85 (55 AU) and 3″.9 (75 AU) on the southwest (SW) side and 2″.33 (45 AU) and 3″.11 (60 AU) on the northeast (NE). This periodic clumping resembles the morphology beyond 500 AU from the star as imaged by *HST*, where surface brightness peaks have been interpreted as evidence for multiple rings in the outer disk (Kalas et al. 2000). Asymmetries similar to those imaged at 10 μm (Lagage & Pantin 1994) are also apparent in Figure 1; the SW side is broader and more elongated. In summary, high-resolution mid-infrared imaging of the inner region of β Pic suggests that a wealth of detail lies just below the limit of spatial resolution. To better identify physically important features, we analyze the image in Figure 1 in two

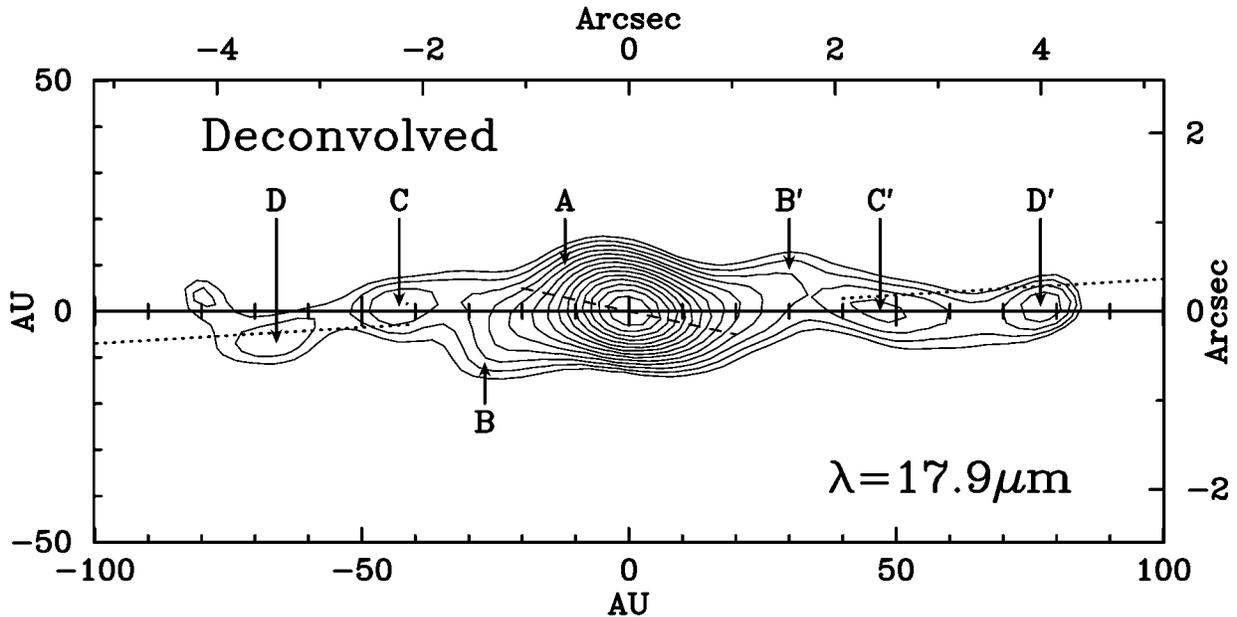

Fig. 2.—Deconvolved image of β Pic at 17.9 μm. The image is oriented as in Fig. 1. Labels A, B, C, D, etc., show positions of possible ring ansae imaged edge-on. Contour levels are the same as in Fig. 1.

ways: Richardson-Lucy deconvolution and Bayesian fitting of a plausible model of the emission.

### 3. MODELING

The image in Figure 1 represents the convolution of the Keck/MIRLIN PSF with the true intensity pattern of mid-infrared emission. We used a Richardson-Lucy algorithm (Richardson 1972; Lucy 1974) to deconvolve the image in Figure 1 with a PSF derived from observations of α Car. The latter were obtained at the same time and in the same region of the sky as β Pic (α Car is only 6° away). Images of α Car taken on all three nights were combined with the same weighting and smoothing scheme as those of β Pic. We restricted our deconvolution to 25 iterations of the Richardson-Lucy algorithm so as to limit the superresolution of the image to a factor of 2.

The deconvolved image of β Pictoris is displayed in Figure 2 with the same contour levels used in Figure 1. The long axis of the core emission (labeled A in Fig. 2) is offset from the plane of the disk as in Figure 1. However, three additional symmetric features are more easily identified. Between 1″ (19 AU) and 1″.5 (29 AU), the long axis of emission is oriented more in the direction of the HST warp (labeled B and B′ in Fig. 2), as indicated by the two-component Gaussian fit. Other isolated peaks are apparent at larger separations and in approximately symmetrical locations on both sides of the star. For convenience, we label these as C (NE, 2″.23, 43 AU), C′ (SW, 2″.44, 47 AU), D (NE, 3″.42, 66 AU), and D′ (SW, 3″.99, 77 AU) in Figure 2. The peaks labeled D and D′ appear to lie along the axis defined by the HST Space Telescope Imaging Spectrograph (STIS) warp at the same distance. Gaps between the emission peaks are located 1″.82 (35 AU) from the star on both sides, at 2″.85 (55 AU) along the NE limb and at 3″.37 (65 AU) in the SW direction.

We speculate that the peaks in Figure 2 constitute the ansae of rings viewed in edge-on projection (e.g., Holland et al. 1998; Koerner et al. 1998). If so, a strikingly complex picture emerges of the inner region of the disk around β Pic. Only one of the putative ring systems is coplanar with the outer disk, and the innermost system is highly inclined relative to all the others. Moreover, eccentricity is strongly indicated for the outermost D ring by offsets in stellocentric distances between mirror-image peaks on either side of the star. Taken together, these features are inconsistent with a continuous inner disk and imply that the depleted inner region inside of β Pic's disk contains four noncoplanar rings, at least one of which is not circular. To better ascertain basic ring properties under this assumption, we vary the properties of a model of four edge-on rings in a Bayesian method of parameter estimation.

Six parameters are required to specify each ring: radius $r$, width $\delta r$, orbital inclination $i$, half-ring thickness $\delta Z$ (height above midplane), face-on optical depth $\tau$, and effective grain size $\lambda_0$. To speed computation, we assume purely circular, edge-on rings that are unresolved with $\delta r = 2$ AU (0″.1). A slightly larger value, $\delta Z = 4$ AU, was adopted for disk thickness, and $\lambda_0$ was set equal to 1 μm (Aitken et al. 1993). Since single-wavelength observations are insensitive to the effective grain size, an error in the latter assumption will have no effect on estimates of ring radii and orientations but will lead to different absolute values of $\tau$. No relative differences in $\tau$ will be affected if all the rings have similar grain sizes (but cf. Weinberger et al. 2002). We will explore ring-dependent values of $\lambda_0$ in a future paper with multiwavelength imaging.

The flux density from an element of ring volume is given by $\tau \epsilon_0 B[T(\lambda_0), \lambda] r \delta r \delta Z$, where $B$ is the Planck function, $\lambda = 17.9$ μm is the observation wavelength, and $\epsilon$ is the emissivity. For $\lambda > \lambda_0$, particles radiate inefficiently, and $\epsilon \sim (\lambda_0/\lambda)$. The grain temperature for particles larger than typical wavelengths of absorbed stellar radiation but smaller than typical wavelengths of their thermal emission is $T = 468(L_*/\lambda_0)^{0.2}(r/1 \text{ AU})^{-0.4}$, where $L_* = 8.7$ $L_\odot$ is the luminosity of β Pic. Only $r_j$, $i_j$, and $\tau_j$ were varied to produce models of each of $j = 1, \ldots, 4$ rings, yielding 12 parameters in all. The flux density for a given model was integrated numerically along the line of sight and convolved with the PSF to generate a simulated image. The result was subtracted from the image displayed in Figure 1 to generate $\chi^2$ in the usual way. The relative probability of each particular model, given the data in Figure 1, was calculated as $e^{-\chi^2}$ and summed over individual parameter values to generate probability distributions for the values of each of the three parameters (see Lay, Carlstrom, & Hills 1997 for a more detailed description of this approach).

The most likely parameter values for the four-ring model are listed in Table 1. They differ slightly from a naive analysis of the original image in Figure 1. In particular, the innermost rings are far more inclined relative to the orbital plane of the outer disk and have radii that are slightly larger than suggested by the peak positions of the ansae. Both effects are what would be expected by convolving a highly resolved ring with a blurring function like the Keck/MIRLIN PSF. These results are only valid under the assumptions of our four-ring model. However, the relative probability of the four-ring model with respect to the best-fit two-ring model is $e^{-\chi_4^2} : e^{-\chi_2^2} \sim 10:1$. Thus, the four-ring model is much more likely and provides a nearly complete account of the morphology of emission in Figure 1. Model fits to estimate brightness asymmetry suggested that the SW ansae of the C and D rings were ~20% brighter than their NE counterparts, a result that agrees with the findings of Pantin et al. (1997).

## 4. DISCUSSION

The images presented here reveal new features in the spatial distribution of dust within 100 AU of β Pic. As evident from Table 1, disk material is relatively depleted at stellocentric distances less than 80 AU. However, some dust is detected to well within 19 AU (1″) of the star where it appears to orbit in a plane misaligned with the outer disk. A ringlike radial structure is suggested by clumping along the long axis of the disk. The clumps can be reproduced with a series of rings with offset orbital inclinations.

The spatial distribution of inner material in the β Pic disk is consistent with properties deduced from earlier measurements (Lagage & Pantin 1994; Pantin et al. 1997). The optical depth of the D ring ($R \sim 80$ AU) is approximately 3 times greater than that of the C ring ($R \sim 50$ AU), confirming the model results of Backman et al. (1992), in which a sharp increase in the radial profile of the optical depth was inferred at 80 AU on the basis of model fitting to infrared flux densities and linear scans. An even greater discontinuity in optical depth exists between the C ring and the B ring ($R \sim 30$ AU), in agreement with an inferred further clearing of the region inside 40 AU (Lagage & Pantin 1994). We also find a brightness asymmetry between the SW and NE ansae, in general agreement with earlier 12 μm imaging (Pantin et al. 1997).

Finally, the outermost D ring ~4″ from the star is in a region studied with HST/STIS imaging (Heap et al. 2000). As identified in Figure 2, it has a radius and angular offset that match the HST warp.

Theoretical explanations for ringlike structure in disks have invoked gravitational interactions with planetary bodies (Liou & Zook 1999; Wyatt et al. 1999; Lubow, Seibert, & Artymowicz 1999; Ozernoy et al. 2000; Wilner et al. 2002), grain evolution and migration under radiation pressure, gas drag, collisional destruction (Klahr & Lin 2001; Takeuchi & Artymowicz 2001), or the gravitational influence of a star passing nearby (Kalas et al. 2000; Kenyon & Bromley 2002). These simulations typically yield only one or two rings and may fail to reproduce the observed anisotropies and ring separations. Models that focus on the influence of a single planet or stellar companion fare somewhat better at producing ring anisotropy and offset orbital inclinations (e.g., Larwood & Papaloizou 1997; Mouillet et al. 1997; Augereau et al. 2001). Our observations suggest that the β Pic system is too complex to be explained by the gravitational influence of a single planet or companion.

TABLE 1
Ring Parameter Estimates[a]

| Ring | R (AU) | i (deg) | τ |
|---|---|---|---|
| A ...... | 14 ± 1 | −32 ± 2 | 5.9 × 10⁻³ |
| B ...... | 28 ± 3 | +25 ± 2 | 2 × 10⁻³ |
| C ...... | 52 ± 2 | −2 ± 2 | 7.7 × 10⁻³ |
| D ...... | 82 ± 2 | +2 ± 1 | 2.3 × 10⁻² |

[a] Rings labeled as in Fig. 2.

A multiple-planet association is supported further by a suggestion of orbital commensurabilities implicit in the radial location of the inner rings. Orbits at 51 and 81 AU (C and D rings) are in 1:2 mean motion resonance, orbits at 14 and 29.1 AU (A and B rings) are in 1:3 mean motion resonance, and those at 29.1 and 51 AU (B and C rings) are in 7:3 mean motion resonance. These features greatly strengthen the case for the existence of a *planetary system* around β Pic. Numerical simulations are sorely needed to test the above interpretations. These take on added importance in the case of any putative planets associated with the A ring since it lies within the habitable zone for β Pic.

We thank John Larwood and David Trilling for useful discussions. Portions of this work were carried out at the Jet Propulsion Laboratory, operated by the California Institute of Technology under a contract with NASA. Data presented herein were obtained at the W. M. Keck Observatory, which is operated as a scientific partnership among the California Institute of Technology, the University of California, and the National Aeronautics and Space Administration. The Observatory was made possible by the generous financial support of the W. M. Keck Foundation. The authors wish also to recognize and acknowledge the very significant cultural role and reverence that the summit of Mauna Kea has always had within the indigenous Hawaiian community. We are most fortunate to have the opportunity to conduct observations from this mountain.